\documentclass[a4paper]{jpconf}
\usepackage{amsmath}
\usepackage{amsfonts}
\usepackage{amssymb}
\usepackage{hyperref}
\usepackage{latexsym}
\usepackage[dvips]{graphicx}
\usepackage{epsf}
\usepackage{bbm}

\newcommand{\be}{\begin{equation}}
\newcommand{\ee}{\end{equation}}
\newcommand{\bea}{\begin{eqnarray}}
\newcommand{\eea}{\end{eqnarray}}
\newcommand{\nn}{\nonumber}

 \let\b=\beta  \let\g=\gamma  
       \let\k=\kappa \let\l=\lambda
\let\m=\mu    \let\n=\nu          \let\r=\rho \let\om=\omega
\let\s=\sigma      
    
    \let\L=\Lambda 
     
  \let\eps=\epsilon

\let\==\equiv

\newcommand{\f}{\frac}

\newcommand{\na}{\nabla}
\newcommand{\w}{\wedge}

\newcommand{\psib}{\overline{\psi}}

\newcommand{\Ref}[1]{(\ref{#1})}

\begin{document}




\title{Perturbative running of the Immirzi parameter\footnote{Based on a talk given by the first author at LOOPS'11, Madrid, May 2011. To appear in the Proceedings.}}

\author{Dario Benedetti${}^1$ and Simone Speziale${}^2$}

\address{${}^1$Max Planck Institute for Gravitational Physics (Albert Einstein Institute), \\
Am M\"{u}hlenberg 1, D-14476 Potsdam, Germany}
\address{${}^2$Centre de Physique Th\'eorique, UMR6207 CNRS-Luminy Case 907, 13288 Marseille, France}

\ead{dario.benedetti@aei.mpg.de, simone.speziale@cpt.univ-mrs.fr}

\begin{abstract}
We report on the renormalization of the Immirzi parameter found through perturbative one-loop calculations.
\end{abstract}


If one takes the tetrad $e_\m^I$ and the Lorentz connection $\om_\m^{IJ}$ as independent variables, the lowest order action for gravity (with zero cosmological constant) has the following form,
\be \label{S1}
S[e,\om] = -\f{1}{\k^2}  \int \left\{\f12 \eps_{IJKL} e^I\w e^J\w F^{KL}[\om] + \f1\g  e_I\w e_J\w F^{IJ}[\om]\right\}.
\ee
The first term is just the Einstein-Cartan action for general relativity. The second term, henceforth Holst term, is, by itself, topological, and does not change the equations of motion in the absence of torsion sources. Its coupling constant $\g$ is known as the Immirzi, or Barbero-Immirzi, parameter in the loop quantum gravity community, where it plays an important role at the non-perturbative level \cite{Rovelli}. This motivates the question of whether any non-trivial quantum mechanical role of $\g$ shows up already in perturbation theory, 
where the gravitational theory \eqref{S1} is non-renormalizable, but can be regarded as an effective field theory. To address the question, in \cite{noi} we studied the perturbative one-loop effective action of \Ref{S1}. 
Here we report on the aspects of our findings most relevant to the renormalization of $\g$ and their implications, and further include explicit plots of the runnings.


To quantize the theory, we used the background-field method and a one-loop perturbative expansion.
We took generic background fields $e$ and $\om$, in particular off-shell, but we assumed invertibility of the tetrad.
This allowed us to decompose $\om=\om(e)+K$, where $\om(e)$ is the unique spin connection solving $d_\om e=0$, and $K$ the contorsion.
Within this framework, the quantization generates an infinite number of terms, that can be packaged in invariants of the background Riemann and contorsion tensors. 
We regularized the theory using the heat-kernel expansion with an UV cut-off $\Lambda_{UV}$. For the divergent part of the one-loop effective action, we found

\be\label{EA}
\Gamma^{\rm div}[e,K] = -\frac{1}{32 \pi^2} \bigg\{ \L_{UV}^4 \int e
 - {\L_{UV}^2} \int e {\cal L}_1  - \ln(\L_{UV}^2/\m^2) \int e  {\cal L}_2 \bigg\},
\ee
with
\bea \label{L1}
{\cal L}_1[e,K] &=& \f{17}{3} R(e) + \f{3\g^2+5}{4\g^2} K_{\m\n\r}K^{\m\n\r} 
 - \f{7\g^2-3}{4\g^2} K^\m{}_{\m\r}K_{\n}{}^{\n\r} + \f{9\g^2-13}{4\g^2} K_{\m\n\r}K^{\n\m\r} \nn\\
 && +
 \f{3}{\g e} \eps^{\m\n\r\s} K_{\m\n}{}^{\l} K_{\r\s\l} +\f{4}{\g e} \eps^{\m\n\r\s} K_{\m\n\r} K^\l{}_{\l\s},
\eea
and ${\cal L}_2$ a very long expression including all possible dimension-four invariants.

In \Ref{L1}, the first $\eps KK$ invariant coincides with the Holst term once $\om=\om(e)+K$ is used.
It is thus the divergent term candidate to renormalize the Immirzi parameter.
However, on-shell $R_{\m\n}=K_{\m\n\r}=0,$
and all the quadratic and logarithmic divergences in \eqref{EA} vanish. Therefore, they can be simply renormalized through field redefinitions.
Only the quartic divergence survives, but it can be reabsorbed, for instance, including also a bare cosmological constant in the action, as in \cite{ChristensenDuff}.
Hence, like in the metric formulations \cite{'tHooft,ChristensenDuff,BuchbinderShapiro}, we find that pure quantum gravity is on-shell finite  at one-loop, and we expect the appearance of non-renormalizable divergences at two loops \cite{GoroffSagnotti}. 

\begin{figure}[h]
\centering \includegraphics[width=7cm]{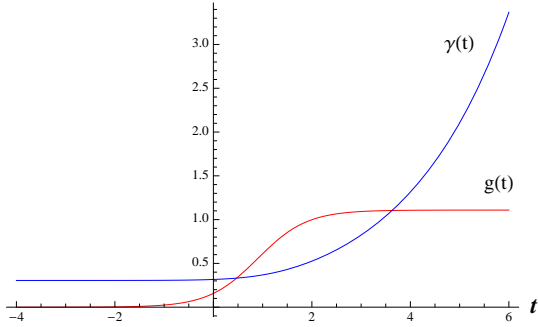}\hspace{2pc}%
\begin{minipage}[b]{6.5cm}\caption{\label{PureRun}\small The running in the pure gravity case, with $t=\ln\mu$ and $\mu$ the energy scale. The running of $g$ is characterized by two plateaus, respectively at the IR Gaussian fixed point, and at the UV non-Gaussian fixed point, with a short transient in between. The running of $\g$ is clearly dictated by the plateaus of $g$, with the first corresponding to a frozen $\g$ and the second to an exponential running.}
\end{minipage}
\end{figure}

The situation can change in the presence of curvature or torsion sources.
As a first step in that direction, we considered a different renormalization scheme, in which the divergences are absorbed whenever possible into a redefinition of the couplings. 
Such a scheme, used for example in \cite{BuchbinderShapiro}, comes closer to the spirit of the calculations done in the context of the asymptotic safety scenario \cite{NiedermaierReuter}, where the running of the traditionally inessential Newton's constant can be motivated by the special role it has in the theory \cite{PercacciPerini}.
With these considerations in mind, we considered an off-shell renormalization condition for $\kappa^2$ and $\g$ obtained from the quadratic divergences \Ref{L1}.
To define non-trivial beta functions from quadratic divergences, we used a non-minimal subtraction ansatz as in  \cite{Robinson, Niedermaier}.
Defining the dimensionless Newton's constant $g\equiv\f{1}{16\pi}\m^2\k^2$, we found the following beta functions,
\be\label{betapure}
\b_g = 
g ( 2 -  \f{17}{3 \pi} g ), \qquad
\b_\g = 
\frac{4}{3 \pi} \g g.
\ee
The beta function for Newton's constant is independent of $\g$,
and it shows an anti-screenig effect and a non-Gaussian fixed point. Although the latter lies outside the realm of perturbation theory, the result is in nice agreement with the asymptotic safety conjecture \cite{NiedermaierReuter}.
The Immirzi parameter has a non-trivial running as well. By inspection,
$\g=0$ and $\g=\infty$ are fixed points, consistently with the results of \cite{DaumReuter}.\footnote{On the other hand, nothing happens at $\g^2=1$. These are the special values for which one is dealing with a formulation of gravity in terms of self-dual variables only.} 
An explicit solution of the flow shows that $\g=\infty$ is UV-attractive. On the other hand, $\gamma=0$ is not the IR asymptote of generic trajectories, because $\g$ stops running as soon as $g$ goes to zero.
In other words, $\{g=0,\gamma\lessgtr 0\}$ is a marginal direction, and the IR limit of $\g$ depends on the initial conditions. 
The situation is depicted and further explained in Fig.~\ref{PureRun}.
%
%

%
%

The result is intriguing, as a UV flow from $\g=0$ to $\g=\infty$ would be nicely consistent with the idea that the metric captures the degrees of freedom of general relativity at low energies, while the connection field becomes more important at high energies, as suggested by loop quantum gravity.
The reason for this is that the $\g\to 0$ limit of \Ref{S1} gives the second-order metric formalism,\footnote{To see this, notice first that the Holst term equals, up to the topological Nieh-Yan invariant, the torsion-squared $T\w T$. In the path integral, $\f{1}{\g} T\w T$ is singular as $\g\to 0$, but it can be rewritten as  $2B\w T + \g B\w B$ using an auxiliary 2-form field $B$. The limit $\g=0$ now yields a Lagrange multiplier enforcing $T=0$, i.e. the second-order theory.} 
whereas for $\g\to\infty$ we obtain Einstein-Cartan theory and the connection is totally  independent.
The running \Ref{betapure} we found gives qualitative support to this picture, however, it is off-shell, thus scheme and gauge dependent. 

An on-shell running of the Immirzi parameter can be obtained adding a source of torsion. 
To do so, we considered the following coupled gravity-fermion system,
\be\label{Scoupled}
S_{\text{coupled}}[e,\psi] = -\int d^4 x \, e 
\left\{\f{1}{\k^2}  R(e) + \f{i}{2} \psib \g^I e_I^\m \na_\m(e) \psi + \f{3\k^2}{128} \Big( \f{\g^2}{\g^2-1} \Big) A_I A^I \right\},
\ee
where $\psi$ is a Majorana spinor, and $A^I=\bar\psi\g^I \psi$ the axial current. This action is obtained from \Ref{S1} minimally coupled to the Dirac action, once the field equations for $\om$ are used.
The one-loop quantization of an action like \Ref{Scoupled} had been previously considered in \cite{Barvinsky:1981rw}. Merging their results with ours, we found for the quadratic divergences
\be \label{1loop-onshell}
{\cal L}_1 = \f{3}{512} \left(28 \f{\g^2}{\g^2-1} -5\right) \k^4 A^2.
\ee
Unlike in the vacuum case, the logarithmic divergences do not vanish on-shell, thus establishing the non-renormalizability of the theory already at one-loop. 
On the other hand, the quadratic divergence \Ref{1loop-onshell} leads to a renormalization of the Immirzi parameter, this time on-shell, with resulting beta function
\be \label{beta-onshell}
\beta_{\g^2} = - (\g^2-1) \f{\m^2 \k^2}{(8\pi)^2} (23\g^2+5).
\ee

\begin{figure}[ht]
\centering  \includegraphics[width=7cm]{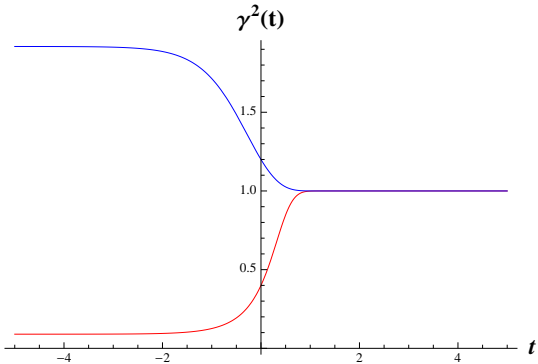}
\hspace{2pc}%
\begin{minipage}[b]{6.5cm}\caption{\label{FermiRun}\small The running of $\g^2$ in the coupled case, with the two sectors $|\g_R| \lessgtr1$. These IR asymptotes depend on the initial condition $\g_0$. Notice that these have to be restricted in order to guarantee that $\g_R(\m)$ remain real for all values of $\m$.
With these restrictions, in the IR limit $\m\to 0$ the Immirzi parameter flows towards a value between $\g_0$ and $|\g_R|=+\infty$, for $\g_0^2>1$, or between $\g_0$ and $\g_R=0$, for $\g_0^2<1$. }
\end{minipage}
\label{fig:vertex}
\end{figure}

We see that neither $\g=0$ nor $\g=\infty$ are stable under renormalization, in contrast with the pure gravity case. There is a fixed point, $\g^2=1$, and it is UV-attractive. However, it corresponds to a divergent coupling for the four-fermion interaction, hence it is out of the range of validity of perturbation theory.
In the IR, the coupling flows to a finite value determined by the initial condition, and the flow depends qualitatively on whether $\g^2$ is larger or smaller than one, with the same UV limit and opposite IR limits. See Fig~\ref{FermiRun} for details.
%
%
%

\noindent Finally, notice that the flow depends explicitly on the external parameter $\m^2 \k^2$. This can be interpreted as the renormalization scale measured in Planck units, as $\k^2$ is inessential and we are not letting it run in the present scheme. 
One can alternatively renormalize $\k^2$ as well, as we did above for pure gravity. In this off-shell scheme, one finds again a $\g$-independent running of Newton's constant, with beta function $\b_g = g_R (2- \tfrac{11}{2\pi}g_R)$, and in 
\eqref{beta-onshell}, ${\k^2\m^2} \equiv 16\pi g_0 \m^2/\m_0^2$ is replaced by a non-trivial running $g_R(\m)$, bounded between $g_R(\m=0)=0$ and $g_R(\m=\infty)={4\pi}/{11}$. This modifies things like the velocity along the flow and the reality bounds on $\g_0$, but not the conclusions about the special points $\g_R^2=0,1,\infty$.

The fact that $\g=0$ is not stable is particularly interesting. If one starts with vanishing bare Immirzi parameter, $\g=0$, the action \Ref{Scoupled} reduces to the second-order Einstein-Hilbert action coupled to fermions, with no four-fermion interaction. However, the latter is nonetheless generated by radiative corrections, see \Ref{1loop-onshell} with $\g=0$.
Namely, the radiative corrections introduce quadratic divergences which are  non-renormalizable in the second-order formalism.
In order to renormalize these divergences one is forced to introduce the four-fermion term in the classical action, that is, one is forced to have a non-vanishing Immirzi parameter.
In this sense, the first order formulation is more suitable to quantize the coupled gravity-fermion system.
Similarly, fine-tuned couplings such as the one proposed in \cite{Mercuri:2006um} do not appear to be stable under radiative corrections.

We also looked at more general actions than \Ref{Scoupled}, which include non-minimal couplings of the fermions.
It can then happen that the dependence on the Immirzi parameter becomes undistinguishable from the non-minimal couplings \cite{Mercuri:2006um, Alexandrov:2008iy}.
In these cases, we found that quadratic divergences non-renormalizable within the Holst action are expected, 
therefore one should include all dimension-two invariants, as in \Ref{L1}, in the bare action.

In conclusion, we showed how the Immirzi parameter gets renormalized by radiative corrections in the perturbative context. 
In the cases considered, the renormalization is driven by quadratic divergences, however a logarithmic contribution to the running likely arises if one includes a cosmological constant term.
The results obtained could serve as a guidance in studies of semiclassical limit of loop quantum gravity, as well as a bridge to the asymptotic safety scenario.

\ack 
We thank Daniel Litim for useful discussions, and we gratefully acknowledge support from the European Science Foundation (ESF) through the activity ``Quantum Geometry and Quantum Gravity''.

\section*{References}



\end{document}